\documentclass[twocolumn]{aastex631}
\usepackage[version=3]{mhchem} 
\begin{document}

\title{Isotopomer-Specific Carbon Isotope Ratio of Complex Organic Molecules in Star-Forming Cores}

\correspondingauthor{Ryota Ichimura}
\email{ryotaichimura.astrolife@gmail.com}

\author[0009-0002-3208-3296]{RYOTA ICHIMURA}
\affiliation{Division of Science, National Astronomical Observatory of Japan, 2-21-1 Osawa, Mitaka, Tokyo 181-8588, Japan}
\affiliation{Department of Astronomical Science, The Graduate University for Advanced Studies, SOKENDAI, 2-21-1 Osawa, Mitaka, Tokyo 181-8588, Japan}

\author[0000-0002-7058-7682]{HIDEKO NOMURA}
\affiliation{Division of Science, National Astronomical Observatory of Japan, 2-21-1 Osawa, Mitaka, Tokyo 181-8588, Japan}
\affiliation{Department of Astronomical Science, The Graduate University for Advanced Studies, SOKENDAI, 2-21-1 Osawa, Mitaka, Tokyo 181-8588, Japan}

\author[0000-0002-2026-8157]{KENJI FURUYA}
\affiliation{RIKEN Cluster for Pioneering Research, 2-1, Hirosawa, Wako-shi, Saitama 351-0198, Japan}

\author[0000-0002-4991-4044]{Tetsuya Hama}
\affiliation{Komaba Institute for Science, The University of Tokyo, Meguro, Tokyo 153-8902, Japan}

\author[0000-0001-5178-3656]{T. J. Millar}
\affiliation{Astrophysics Research Centre, School of Mathematics and Physics, Queen’s University Belfast, University Road, Belfast BT7 1NN, UK}

\begin{abstract}
The recent observation of complex organic molecules (COMs) in interstellar ices by the James Webb Space Telescope (JWST), along with previous gas-phase detections, underscores the importance of grain surface and ice mantle chemistry in the synthesis of COMs.
In this study, we investigate the formation and carbon isotope fractionation of COMs by constructing a new astrochemical reaction network that distinguishes the position of $^{13}$C within species (e.g., H$^{13}$COOCH$_3$ and HCOO$^{13}$CH$_3$ are distinguished).
We take into account the position of $^{13}$C in each species in gas and solid phase chemistry. This new model allows us to resolve isotopomer-specific $^{12}$C/$^{13}$C ratios of COMs formed in the star-forming cores. We consider thermal diffusion-driven radical-radical reactions on the ice surface and non-thermal radiolysis chemistry in the bulk (surface + mantle) ice.
We find that carbon isotope fractionation of the functional groups in COMs appears through both non-thermal radiolysis in cold environments and thermal diffusion in warm environments, depending on the COMs.
In particular, COMs containing methyl groups show isotopomer differences in $^{12}$C/$^{13}$C ratios that reflect their formation pathways and environments.
These isotopomer-resolved fractionation patterns provide a diagnostic tool to probe the origins of COMs in star-forming cores.
Our results suggest that future comparisons between high-sensitivity isotopic observations and isotopomer-specific models will be helpful for constraining the relative contributions of thermal and non-thermal formation processes of COMs.
\end{abstract}


\section{Introduction}
The origin of complex organic molecules (COMs), typically defined as organic molecules with six or more atoms, has been a subject of investigation in astrochemistry {over the past two decades}\citep{2009ARA&A..47..427H}.
They have been detected in various environments in star-forming regions, and recent observations by the James Webb Space Telescope (JWST) have confirmed the presence of functional groups associated with COMs in interstellar ices around protostars \citep{2023NatAs...7..431M}. 
Since interstellar ices on dust grains are the fundamental building blocks of planetesimals—precursors of comets and asteroids—the chemical composition of these ices directly influences the inventory of organic matter delivered to planetary systems.
Understanding the formation pathways of COMs is therefore critical not only for tracing the chemical evolution from molecular clouds to planets, but also for evaluating their potential relevance to prebiotic chemistry in the early Solar System \citep{caselli2012our, 2023ASPC..534..379C}.

While numerous models have been proposed to account for COMs formation, many questions remain unresolved, especially regarding which types of ice chemistry dominate under different physical conditions.
In cold environments, such as molecular clouds, COMs are proposed to be formed through the diffusion reaction between radicals among the weakly bound sites on the grain surface \citep{2024ApJ...974..115F, 2024A&A...688A.150M} or non-diffusive grain surface and ice mantle chemistry, such as the Eley-Rideal (ER) mechanism \citep{ruaud2015modelling}, three-body reaction mechanism \citep{2020ApJS..249...26J, 2022ApJS..259....1G}, or cosmic-ray-induced radiolysis \citep{2018ApJ...861...20S}. In a warm environment, thermal diffusion reaction between radicals is thought to form COMs on the grain surface \citep{2006A&A...457..927G}. 
{In addition to solid-state processes, several studies have also shown that COMs can form through gas-phase chemistry under both cold and warm conditions \citep{2015MNRAS.449L..16B, 2023ApJ...944..208B, 2025MNRAS.537.3861G}}.
Despite these advances, it remains challenging to determine the relative importance of these pathways for each COMs across various astrophysical environments.
{Some recent studies have detected the same COMs (e.g., C$_2$H$_5$OH) both in interstellar ices and in the gas phase at higher temperatures \citep{2023NatAs...7..431M, 2018Jorgensen}, but it remains unclear where these COMs are primarily formed.
The same molecule can originate from different pathways—such as grain-surface reactions, ice-mantle chemistry, or gas-phase processes—and the product species alone do not reveal their formation history.}

Isotope fractionation offers a diagnostic tool for probing molecular formation histories \citep{2023Nomura}. 
Carbon isotope ratios of species that deviate from the average local interstellar medium (ISM) value ($^{12}$C/$^{13}$C $\sim$69) \citep{2005ApJ...634.1126M} have been observed in various molecules across a range of astrophysical environments.
For example, observations toward molecular clouds have reported variations in the $^{12}$C/$^{13}$C ratios among molecules such as CO and CN \citep{1993ApJ...408..539L, 2005ApJ...634.1126M}.
These molecular-scale isotopic differences are thought to reflect distinct formation histories and chemical environments. 

Additionally, isotopomers where the $^{13}$C substitution occurs at different atomic positions have also been detected.
{For instance, observations in star-forming regions show that carbon isotopomers of various species exhibit different abundances—for example, $^{13}$CCH and C$^{13}$CH in simple hydrocarbons \citep{2010A&A...512A..31S,2019A&A...625A.147A}, $^{13}$C-substituted HC$_3$N formed mainly in the gas phase \citep{1998A&A...329.1156T,2024A&A...682L..12T}, and thermally desorbed COMs such as $^{13}$CH$_3$CHO and CH$_3$$^{13}$CHO observed in V883 Ori} \citep{2024AJ....167...66Y}.
{These results indicate that carbon atoms at different positions within a molecule can exhibit distinct $^{13}$C enrichments, demonstrating that isotopomer-specific fractionation can occur through both gas-phase and ice-surface formation pathways.}
{In particular, isotopomer-specific signatures detected in species such as HC$_3$N indicate that their $^{13}$C distribution reflects gas-phase formation pathways.
Furthermore, recent studies have emphasized that in low-density environments ($\sim$ 10$^3$ cm$^{-3}$), COMs formation on ices becomes inefficient, and gas-phase reactions may play a more important role in determining isotopomer-specific carbon fractionation \citep{2023ApJ...944..208B}.
These findings suggest that a comprehensive interpretation of $^{12}$C/$^{13}$C ratios and isotopomer differences requires treating both gas-phase and ice-phase processes consistently.
In this study, however, our primary focus is in dense prestellar and protostellar environments, where ice-phase pathways dominate COMs formation.}

Moreover, isotopic measurements of meteoritic organic compounds—including amino acids and other prebiotically relevant molecules—have revealed that the $^{13}$C content can differ between functional groups within a single molecule \citep{zeichner2023position}.
These differences are believed to encode information about the molecular formation processes and environments, pointing to a position-specific imprint of isotope fractionation.
However, previous astrochemical models for $^{13}$C fractionation in COMs have generally assumed full scrambling of carbon atoms within molecules, treating all positions of carbon atoms as equivalent \citep{2024ApJ...970...55I}. While this simplification facilitates the mechanical construction of chemical networks, it overlooks the possibility that isotope effects may vary between atomic positions.

To address this limitation, we develop an astrochemical model that explicitly tracks the position of $^{13}$C atoms throughout reaction sequences, allowing for the evaluation of isotopomer-specific carbon fractionation in COMs under astrophysical conditions.
By using cheminformatics tools, we systematically track the position of carbon atoms across chemical reactions. With our new chemical network, we evaluate how the resulting isotopomer ratios reflect the physical environment and chemical history of COMs in star-forming regions.
The rest of this paper is structured as follows: Section 2 describes the construction of our position-conserved reaction network and the physical and chemical model used in the simulations. Section 3 presents the results of our calculations during the static and collapse phases. In Section 4, we discuss the implications of our findings, including comparisons to the full-scrambling model and observed isotopomer ratios. Finally, we summarize our findings in Section 5. 
To avoid ambiguity in terminology, we clarify our usage of carbon isotope and isotopomer ratios.
The ``carbon isotope ratio" refers to the abundance ratio between a molecule and its singly $^{13}$C-substituted isotopologue, that is, the main isotopologue ($^{12}$CX) to each $^{13}$C-bearing isotopologue ($^{13}$CX), where only one $^{12}$C is replaced by $^{13}$C. In molecules with multiple carbon atoms, this ratio is defined for each specific carbon position, corresponding to structurally distinct isotopomers such as $^{13}$CH$_3$CHO and CH$_3$$^{13}$CHO. 
For symmetric molecules in which multiple carbon atoms are chemically and spectroscopically indistinguishable, the carbon isotope ratio is derived from the abundance ratio between the main isotopologue and the sum of all singly $^{13}$C-substituted isotopologues, multiplied by a statistical factor equal to the number of equivalent carbon positions.
{For example, in the case of the C$_3$ molecule, the $^{13}$C-substituted isotopomers $^{13}$CCC and CC$^{13}$C are chemically equivalent owing to the molecular symmetry. In our model, these two isotopomers are treated as identical and represented by CC$^{13}$C. Accordingly, when evaluating the carbon isotope ratio, we compute
$^{12}$C/$^{13}$C (terminal carbon) of C$_3$ = ([C$_3$]/[CC$^{13}$C])$\times$2,
where the factor of 2 accounts for the two equivalent terminal carbon positions.}
On the other hand, the central carbon–substituted isotopomer C$^{13}$CC is not equivalent to the terminal ones and can be treated separately. Thus, the carbon isotope ratio for the central position is simply given by: $^{12}$C/$^{13}$C of central carbon = [C$_3$]/[C$^{13}$CC].
This correction allows estimation of the $^{12}$C/$^{13}$C ratio per carbon position. A deviation of the carbon isotope ratio from the elemental abundance ratio of [$^{12}$C/$^{13}$C] in the local ISM (e.g., 69) is referred to as ``isotope fractionation".
The ``isotopomer ratio" refers specifically to the abundance ratio between two distinct singly $^{13}$C-substituted isotopomers of a molecule with more than one carbon atom (e.g., $^{13}$CH$_3$CHO / CH$_3$$^{13}$CHO). A deviation of this ratio from unity indicates ``isotopomer difference".

\section{Model Description} \label{sec:model}
\subsection{Overview}
\begin{figure*}[htbp]
\centering
\hspace*{-0.3cm}
\includegraphics[height=0.2\hsize]{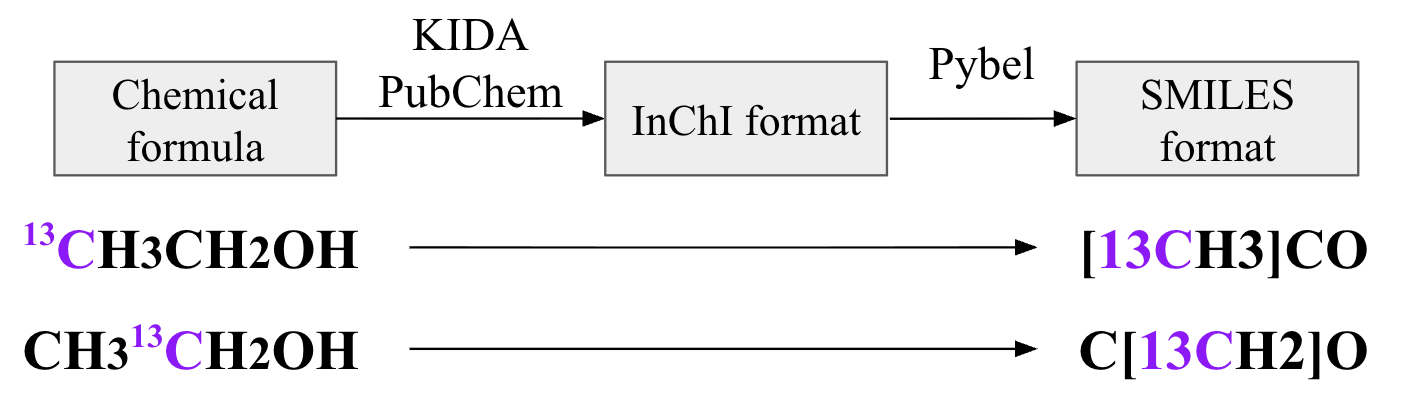}
\vspace{1cm}
  \centering

  \newlength{\imgheight}
  \setlength{\imgheight}{3.5cm}

  \begin{minipage}[t]{0.28\linewidth}
    \centering
    \raisebox{1.05cm}{\includegraphics[height=2.5cm]{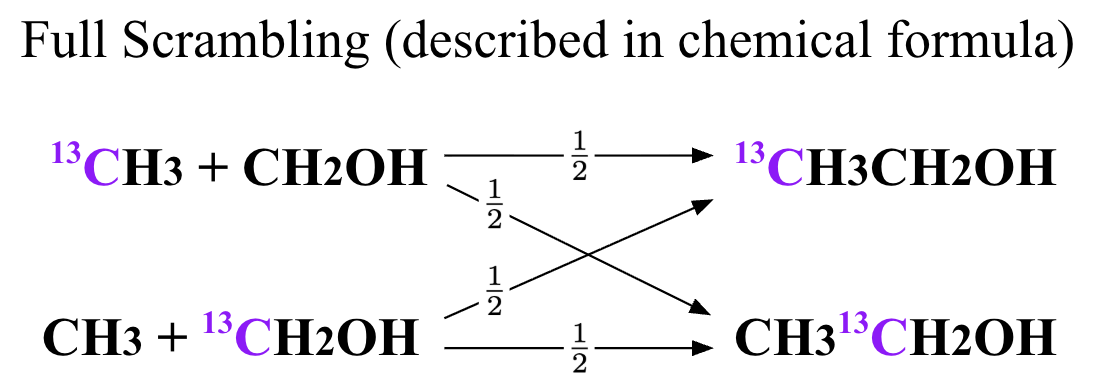}} 
  \end{minipage}
  \hfill
  \begin{minipage}[t]{0.6\linewidth}
    \raggedleft
    \includegraphics[height=\imgheight]{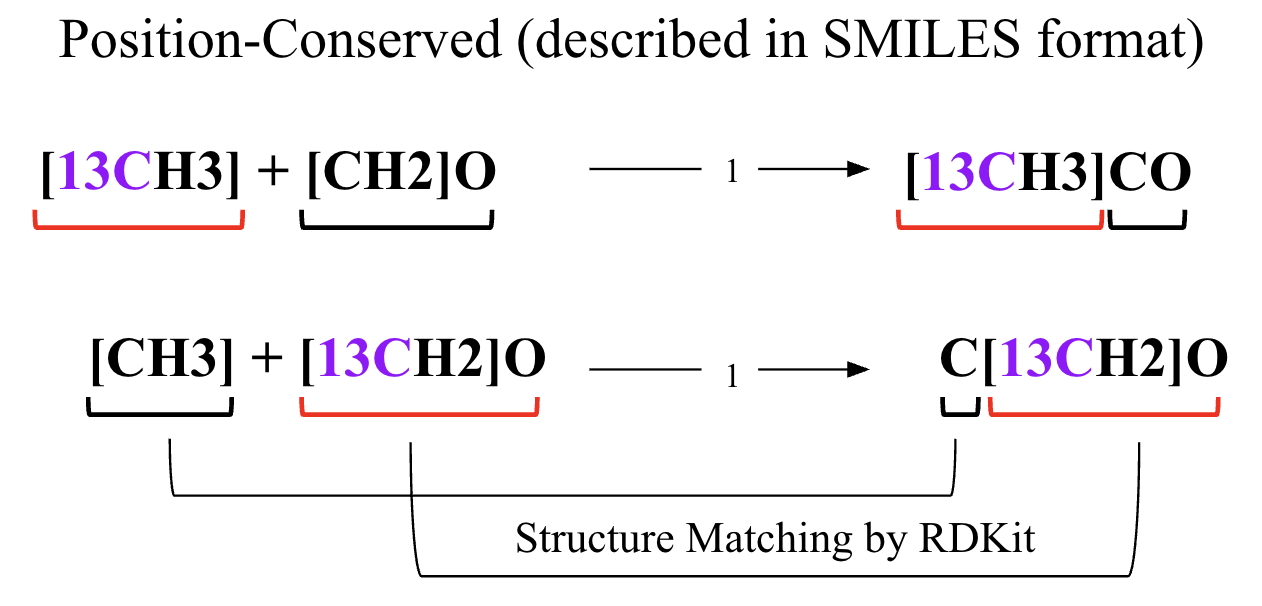}
  \end{minipage}

  \caption{Conversion of chemical formula to SMILES format representation (upper). Comparison between the full scrambling (lower left) and position-conserved (lower right) models taking the reaction for the formation of ethanol (C$_2$H$_5$OH), CH$_3$ + CH$_2$OH $\rightarrow$ C$_2$H$_5$OH as an example.}
  \label{fig: composite}
\end{figure*}

A new chemical reaction network has been developed to describe carbon isotope fractionation among isotopomers. This network is based on \citet{garrod2013three}, with modifications that include $^{13}$C isotope chemistry, non-thermal radiolysis chemistry in the bulk ice, and several other minor updates. Details of the main modifications are provided in the following sections.
These modifications include the addition of $\sim$1600 new isotopically labeled species and $\sim$45500 reactions involving $^{13}$C-bearing molecules and non-thermal radiolysis reactions. As a result, our complete reaction network consists of approximately $\sim$2900 species across the gas, ice surface, and ice mantle phases, and $\sim$59000 reactions covering both gas-phase and ice-phase chemistry.
We solve the updated chemical reaction network using a gas-ice astrochemical code based on the rate equation approach (Rokko code; \citet{2015Furuya}).
We adopt the three-phase gas-ice chemistry model developed by \citet{h&h1993}, consisting of three components: the gas phase, the ice surface, and the bulk ice mantle. 
Following \citet{2013ApJ...762...86V}, we treat the top four monolayers of the ice mantle as chemically active.
Simulation of the carbon isotope chemistry is performed under physical conditions evolving from a prestellar core to a protostellar core as described by \citet{2024ApJ...970...55I}.

\subsection{Carbon Isotope Chemistry}
{Carbon isotope exchange reactions in the gas phase are included to reproduce isotopic fractionation among the main carbon reservoirs (C$^+$, C, and CO).
These reactions are primarily driven by differences in the zero-point vibrational energies between isotopologues, which make the formation of $^{13}$C-enriched species energetically favorable at low temperatures \citep{watson1976,langer1984,2015Roueff}.
The $^{13}$C fractionation established in the gas phase is transferred to the ice via adsorption and desorption, where it is inherited by the molecular precursors of COMs.
Consequently, these exchange reactions regulate the $^{13}$C distribution among small carbon-bearing species and influence the resulting $^{12}$C/$^{13}$C ratios of COMs formed on icy grains.
As demonstrated in our previous study \citep{2024ApJ...970...55I}, exchange involving small species such as C, C$^+$, and CO determines the isotopic compositions inherited by key COMs precursors (e.g., CH$_4$ and CH$_3$OH ices), which in turn shape the $^{12}$C/$^{13}$C ratios of COMs.
In the present model, only isotope-exchange reactions involving a single $^{13}$C atom are considered.
We adopt 25 carbon isotope exchange reaction pairs compiled by \citet{colzi2020,loison2020,sipila2023combined}.}
We extend the \citet{garrod2013three} reaction network by explicitly considering the positional distribution of $^{13}$C atoms in isotopologues containing {up to three carbon atoms} (e.g., $^{13}$CH$_3$COCH$_3$ and CH$_3$$^{13}$COCH$_3$ for CH$_3$COCH$_3$, acetone).
Although the full network includes species with up to ten carbon atoms, this isotopomer-specific treatment is restricted to smaller molecules for computational feasibility and because observational constraints are mostly available for such species.
In gas-phase chemistry, we assume that all reactions involving $^{13}$C, except for proton transfer and charge transfer, proceed through full scrambling as assumed in \citet{colzi2020}. In full scrambling, the reactants form an intermediate complex that persists long enough for the constituent atoms to undergo complete rearrangement \citep{gerlich1992experimental}. Full scrambling involving $^{13}$C isotope species leads to two outcomes (see the left of Figure \ref{fig: composite}). One corresponds to the position of the $^{13}$C atom being retained from a specific reactant into the product, while the other corresponds to the interchange of the $^{13}$C atom between reactants. The branching ratios are constructed based on statistical weights, which reflect the number of carbon atoms and their possible arrangements in the intermediate complex. Specifically, we calculate the branching ratios by multiplying simple probabilities associated with the possible positions of the $^{13}$C atoms in the products. For example, in a reaction involving two carbon atoms, the probability of forming each isotopomer is proportional to the number of indistinguishable ways to assign $^{13}$C to a given carbon position, assuming full scrambling. 
For proton transfer and charge transfer reactions, it is assumed that no atoms, including carbon atoms, within the reactants undergo interchange during the chemical process.

{We first examine the potential reactivity of $^{12}$C and $^{13}$C in various ice-phase reactions (e.g., hydrogen abstraction, hydrogenation, radical–radical, and photodissociation reactions). 
Based on this discussion, we assume that the interchange of carbon atoms ($^{12}$C and $^{13}$C) does not occur in ice-phase reactions, and accordingly construct a reaction network that conserves the position of each carbon atom.}
Based on the theoretical study \citep{2019JPCA..123.9061C}, during hydrogen abstraction from closed-shell molecules such as CH$_3$OH, the carbon atom transiently shifts toward the departing hydrogen atom along the reaction coordinate, but ultimately returns to its original position within the molecular framework. Therefore, we neglect the interchange of carbon atoms through hydrogen abstraction reactions. 
We also neglect the interchange of carbon atoms during hydrogenation reactions. A pre-reactive minimum or a van der Waals complex formed in these reactions does not evolve into an intermediate complex or a transition state where the atomic interchange occurs. Instead, the product is considered to result from hydrogen tunneling from bound metastable states \citep{meisner2016atom}.
Radical–radical reactions in the ice generally have low activation barriers, and their transition states typically involve only the translation and/or rotation of the radicals, thereby preventing the interchange of carbon atoms \citep{2022ApJS..259...39E}. Furthermore, the chemical networks of \citet{garrod2013three} and \citet{2008ApJ...682..283G}, which form the foundation of our model, assume that no intramolecular rearrangement occurs during radical–radical reactions and that the structures of the products directly reflect those of the reacting radicals {in the ice phase}. Consistent with these theoretical and modeling considerations, our $^{13}$C chemical network does not allow for carbon scrambling in radical–radical reactions, thereby preserving the original positions of $^{13}$C within the molecules.  
For example, when each reactant has a carbon atom, we assume that the product inherits positions of $^{12}$C and $^{13}$C in a functional group (e.g., CH$_3$ and CHO) from each {ice reactant}.
For example, in a radical-radical reaction for the formation of acetaldehyde (CH$_3$CHO):
\begin{equation}
    \label{eq:}
    \rm{CH_3 + HCO \rightarrow \ CH_3CHO,}
\end{equation}
the `HCO' in the reactant and the formyl group `CHO' of a product are common structures. When a $^{13}$C atom is present in the `HCO' of the reactant, it is conserved in the `CHO' of the product (Eq.(\ref{eq:2})). Similarly, when a $^{13}$C atom is present in the `CH$_3$' of the reactant, it remains in the methyl group `CH$_3$' of the product (Eq.(\ref{eq:3})):
\begin{equation}
    \label{eq:2}
    \rm{CH_3 + H^{13}CO \rightarrow CH_3^{13}CHO,}
\end{equation}

\begin{equation}
    \label{eq:3}
    \rm{^{13}CH_3 + HCO \rightarrow \ ^{13}CH_3CHO.}
\end{equation}

Dissociation processes within the ice, such as photodissociation or recombination, may cleave only a single chemical bond within a molecule, so the interchange of carbon atoms may not occur. 

We developed a \textsc{Python} script that automatically incorporates $^{13}$C isotopologues and $^{13}$C position-conserved reactions into a chemical reaction network composed of molecular formulae. 
Conventional molecular formulae are insufficient for mechanically constructing such a chemical reaction network with carbon atom position-conserved reactions, because they do not specify which carbon atom, belonging to which functional group, is substituted with $^{13}$C. 
Instead, we utilize the International Chemical Identifier (InChI) and the Simplified Molecular Input Line Entry System (SMILES) as molecular string grammars (see Figure \ref{fig: composite}). InChI and SMILES represent the structure of molecules in formats interpretable by both humans and computers, provide a unique representation for each molecular structure, and can uniquely distinguish between structural isomers that cannot be differentiated by conventional molecular formulae \citep{2012OBoyle, fried2023implementation, 2023ApJ...959..108S}. 
Firstly, we acquire the InChI format of species in \citet{garrod2013three} network from the \textsc{KIDA} (\url{https://kida.astrochem-tools.org/}) and \textsc{PubChem} (\url{https://pubchem.ncbi.nlm.nih.gov/}) databases. InChI strings are converted to SMILES format via \textsc{Pybel} \citep{o2008pybel}, which provides \textsc{Python} access to the \textsc{OpenBabel} library (\url{https://openbabel.org/index.html}). SMILES is a machine-readable notation for molecules, widely used in cheminformatics, a field that applies computational tools to chemical data analysis. We use the open-source toolkit `RDKit' (\url{https://www.rdkit.org/}) to do cheminformatics analysis. The Substructure matching module in RDKit is used to identify common functional groups in reactant and product molecules and track the position of $^{13}$C from reactants to products (see the right of Figure \ref{fig: composite}).
The CanonicalRankAtoms module is used to identify indistinguishable carbon atoms within a symmetric molecule. 
This module assigns a number `rank' to each atom within the species. Atoms with the same rank are assigned to identical atoms in the species.  
If there are two or more $^{12}$C atoms with the same rank in the species, the species in which one of them is replaced by $^{13}$C are considered to be the same species. For example, the dimethyl ether (CH$_3$OCH$_3$) has two indistinguishable carbon atoms, meaning that its isotopologues $^{13}$CH$_3$OCH$_3$ and CH$_3$O$^{13}$CH$_3$ are equivalent. Therefore, in the radical-radical reaction, CH$_3$ + CH$_3$O $\rightarrow$ CH$_3$OCH$_3$,
even if a $^{13}$C atom is present in either reactant, the products of `$^{13}$CH$_3$ + CH$_3$O' and `CH$_3$ + $^{13}$CH$_3$O' are identical.




\subsection{Cosmic-ray-induced Non-thermal Reaction}
We adopt the cosmic-ray-induced radiolysis model as non-thermal chemistry for COMs formation in the ice-phase \citep{2018ApJ...861...20S}. {Cosmic rays}, which include high-energy protons, induce electronic excitation in ice molecules, leading to the formation of ``suprathermal" species. Although suprathermal species have a much shorter lifetime ($\ll$1s) due to the rapid quenching, these suprathermal species are assumed to overcome activation energy barriers and react with neighboring molecules and produce COMs.

In our model, radiolysis processes are assumed to occur both on the surfaces of dust grains and within the bulk ice mantle, following \citet{2018ApJ...861...20S}. Specifically, we adopt the framework of radiolysis-induced generation of suprathermal species as listed in Table 4 of \citet{2018ApJ...861...20S}, while the subsequent reactions between these suprathermal species and neighboring stable molecules are modeled by reusing analogous ice-phase reactions already included in the chemical network of \citet{garrod2013three}.

\subsection{Other Modifications and Parameters}
In addition to the major modifications, we also implement several other modifications and assumptions in our chemical network.
While radical–radical reactions such as CH$_3$ + HCO → CH$_3$CHO are generally considered to be barrierless—as is the case in the gas phase, supported by experimental and theoretical studies \citep{callear1990formation,2017CPL...685..165K}—the presence of surrounding water ice may introduce intermolecular interactions that hinder direct recombination. Following the quantum chemical calculations of \citet{enrique2019reactivity}, we adopt effective energy barriers of 800 K and 1200 K for CH$_3$CHO formation and for the competing hydrogen abstraction pathway CH$_3$ + HCO → CH$_4$ + CO, respectively. These values reflect the energy cost of overcoming interactions with the surrounding ice matrix, rather than intrinsic activation barriers of the reactions themselves.
{We note that \citet{enrique2022quantum} slightly updated the activation energies to 5.5 kJ mol$^{-1}$ and 7.2 kJ mol$^{-1}$, respectively; however, converting these values to Kelvin (approximately 660 K and 870 K) yields comparable barriers, and adopting them does not affect our model results.}
In our model, we also assume that these reactions proceed without carbon atom scrambling. This assumption is based on the fact that the intrinsic reactions are effectively barrierless and that, once the surrounding ice interactions are overcome, the transition state involves only the translation and/or rotation of the radicals. As is typical for radical–radical reactions, no rearrangement of the carbon backbone is expected, and the original position of each $^{13}$C atom within the molecular structure is preserved.
A fraction of the adsorbed carbon undergoes direct Eley–Rideal (ER) reactions with surface H$_2$O and CO ice, following the approach of \citet{2024ApJ...970...55I}, while the remainder becomes trapped in either physisorbed or chemisorbed states. Although physisorbed atomic carbon may retain some mobility, as discussed by \citet{2023Tsuge,tsuge2024methane}, we adopt a binding energy of 10,000 K—consistent with chemisorption values from \citet{wakelam2017binding}—to represent its interaction with the ice, thereby simplifying its treatment in the reaction network.

\begin{table}[t!]
\centering
\caption{Initial abundances with respect to hydrogen nuclei.}
\label{tab:initial_abundances}
\begin{tabular}{lc}
\hline\hline
{Species} & {Abundance ($n_i/n_{\mathrm{H}}$)} \\
\hline
H$_2$   & $5.00\times10^{-1}$ \\
He      & $9.75\times10^{-2}$ \\
$^{12}$C$^{+}$ & $7.86\times10^{-5}$ \\
$^{13}$C$^{+}$ & $1.14\times10^{-6}$ \\
N       & $2.47\times10^{-5}$ \\
O       & $1.80\times10^{-4}$ \\
Si$^{+}$ & $9.74\times10^{-9}$ \\
S$^{+}$  & $9.14\times10^{-8}$ \\
Fe$^{+}$ & $2.74\times10^{-9}$ \\
Na$^{+}$ & $2.25\times10^{-8}$ \\
Mg$^{+}$ & $1.09\times10^{-8}$ \\
Cl$^{+}$ & $2.16\times10^{-10}$ \\
P$^{+}$  & $1.00\times10^{-9}$ \\
\hline
\end{tabular}
\end{table}

We adopt the low-metal elemental abundances listed in Table~\ref{tab:initial_abundances} from \citet{aikawa2001}, along with the elemental $^{12}$C/$^{13}$C ratio representative of the local ISM ($^{12}$C/$^{13}$C = 69) \citep{2005ApJ...634.1126M}. 
The dust grain is spherical with a 0.1 {\textmu}m radius with the material density of 2.5 g/cm$^3${\citep{2003ARA&A..41..241D}}. {The dust-to-gas mass ratio is set to 0.01 \citep{2014A&A...566A..45L}.} The cosmic-ray ionization rate of H$_2$ is set to be 1.3$\times$10$^{-17}$ s$^{-1}$ \citep{1998T&H}. 
{Ice species are assumed to have a single binding energy based on \citet{garrod2013three}}. The diffusion energy is set to 40 \% of the binding energy for most species, except for atomic hydrogen.
We set the diffusion energy of H atoms to 330 K, which corresponds to the energy barrier for the deep potential sites \citep{2012ApJ...757..185H}. 
To prevent nitrogen atoms from diffusing more easily than H atoms, we assume a diffusion energy of 500 K for atomic nitrogen.
In the ice mantle, the binding energies of the species, except H and H$_{2}$, are the same as water ice (5700 K). The diffusion energy of species is set to 80 \% of the binding energy.
In this paper, we define ``abundance" as the fractional abundance of species to hydrogen nuclei.

\begin{figure}[ht!]
\centering
\includegraphics[height=0.5\hsize]
{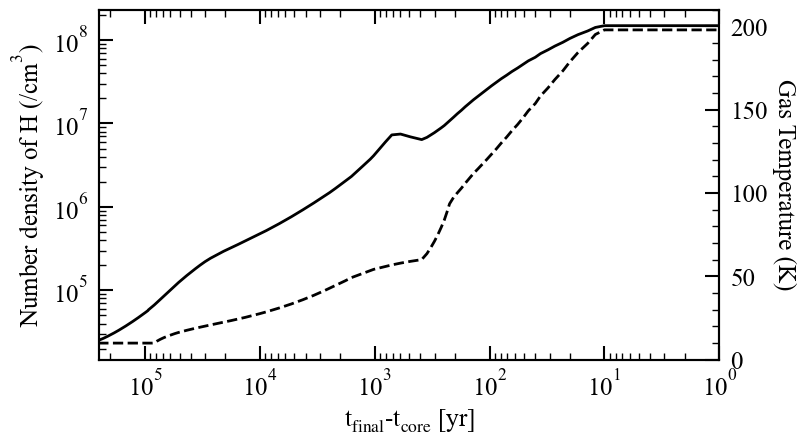}
\caption{Temporal variation of the number density of hydrogen nuclei (solid line) and gas temperature (dashed line) of a fluid parcel along a streamline in a gravitationally collapsing core.
\label{fig: physicalmodel}}
\end{figure}

We calculate the evolution of molecular abundances with a chemical reaction network along a single fluid parcel as done in \citet{aikawa2020}.
The simulation begins with a static prestellar core phase, characterized by a temperature of 10 K, a hydrogen number density of 2.28$\times$10$^{4}$ cm$^{-3}$ and a visual extinction of 4.5 mag.
During this phase, the core is assumed to be in hydrostatic equilibrium with constant physical conditions maintained for 1$\times$10$^6$ years.
{For the subsequent collapse phase, we use the result of a one-dimensional radiation hydrodynamics simulation by \citet{m&i2000}, adopting their {hydrostatic-core initial condition} rather than the homogeneous model. 
This choice provides a well-defined quasi-static configuration suitable for coupling with time-dependent chemistry. 
The collapse begins at $t=0$, and a protostar forms after $2.5\times10^{5}$ years. 
The model then tracks the physical evolution of the system for an additional $9.3\times10^{4}$ years, where the moment of protostar formation is defined as $t_\mathrm{core}=0$ and the final simulation time is $t_\mathrm{core}=t_\mathrm{final}=9.3\times10^{4}$ years. 
Figure \ref{fig: physicalmodel} shows the temporal variation of the number density of hydrogen nuclei and gas temperature in a fluid parcel initially located at 4000 au. 
The {density and temperature profiles} are obtained by interpolating the radial distributions in the \citet{m&i2000} model and converting them into temporal evolution, following the procedure of \citet{aikawa2008}. 
To highlight the rapid changes near the final stage, we adopt a logarithmic scale for $t_\mathrm{final}-t_\mathrm{core}$ on the horizontal axis.}
{The model assumes that the dust temperature is equal to the gas temperature throughout the simulation.}
Our model excludes $^{13}$C fractionation via isotope-selective photodissociation, as the dense regions we simulate render photoreactions negligible.

\section{Results} \label{sec: results}

\subsection{Static Phase} \label{subsec: static}

\begin{figure}[ht!]
\centering
\includegraphics[height=1.3\hsize]{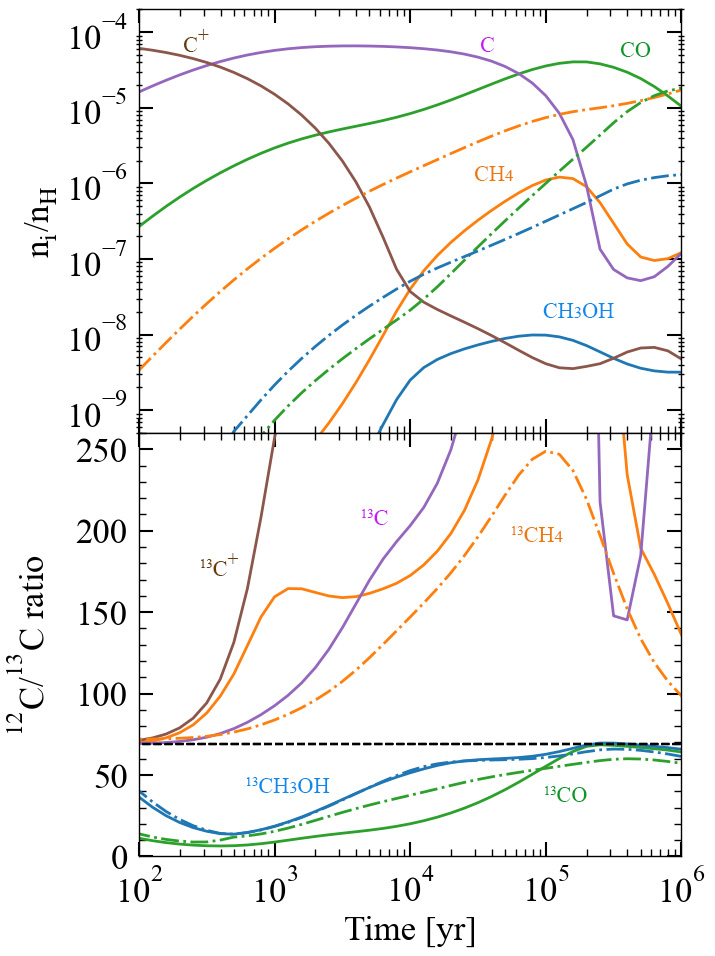}
\caption{Temporal variation of the molecular abundances and $^{12}$C/$^{13}$C ratios for gaseous species (solid lines), the bulk (surface + mantle) ice (dash-dot lines) during the static phase. The horizontal black dashed line represents the elemental $^{12}$C/$^{13}$C ratio.
\label{fig: dominantC_static}}
\end{figure}


\begin{figure*}[t!]
\centering
\includegraphics[height=0.27\hsize]{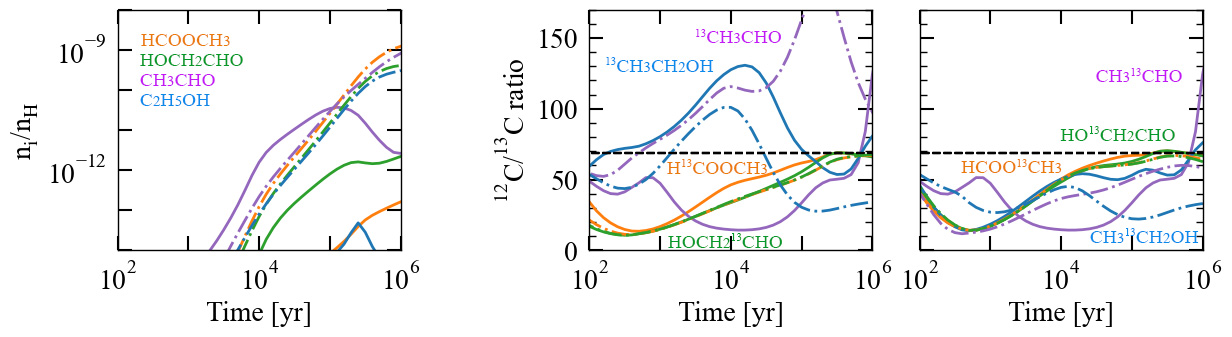}
\hspace{0.3cm}
\includegraphics[height=0.278\hsize]{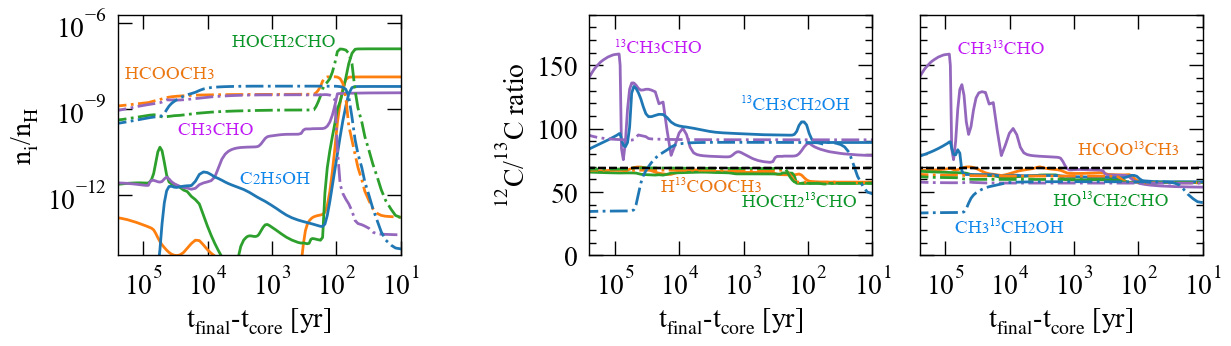}
\caption{
Temporal variation of the abundances (left panels) and $^{12}$C/$^{13}$C ratios (middle and right panels) of selected COMs in the gas phase (solid lines) and in the bulk (surface + mantle) ice (dash-dotted lines) during the static phase (upper panels) and the collapse phase (lower panels). 
The middle panels show $^{12}$C/$^{13}$C ratios for H$^{13}$COOCH$_{3}$, HOCH$_{2}$$^{13}$CHO, $^{13}$CH$_{3}$CHO, and $^{13}$CH$_{3}$CH$_{2}$OH, while the right panels show those for HCOO$^{13}$CH$_{3}$, HO$^{13}$CH$_{2}$CHO, CH$_{3}$$^{13}$CHO, and CH$_{3}$$^{13}$CH$_{2}$OH. 
The vertical axis is shared between the middle and right panels, and the horizontal black dashed line represents the elemental $^{12}$C/$^{13}$C ratio.
}
\label{fig: COMs_static_collapse}
\end{figure*}

Figure~\ref{fig: dominantC_static} shows the temporal evolution of the abundances and $^{12}$C/$^{13}$C ratios for dominant carbon-bearing species in the gas and ice during the static phase. As established in previous studies \citep{furuya2011,colzi2020,loison2020,2024ApJ...970...55I}, carbon isotope exchange reaction
\begin{equation}
    \label{eq:1}
    \rm{^{13}C^{+}+{}^{12}CO\rightleftharpoons{}^{12}C^{+}+{}^{13}CO + 35 K, }
\end{equation}
leads to $^{13}$C enrichment in CO, while C and C$^+$ become $^{13}$C-poor. This fractionation pattern is established within $\sim$10$^3$ years and gradually relaxes toward the local ISM value ($^{12}$C/$^{13}$C = 69) after 10$^5$ years as carbon becomes sequestered in CO {as also reported in \citet{2024ApJ...970...55I}}. These gas-phase fractionation is transferred to the isotope composition of molecules formed on grains via adsorption and surface reactions.

In our model, COMs predominantly form in cold environments (10 K) during the static phase via two-body reactions with suprathermal species induced by radiolysis in the ice mantle.
Upper panels of Figure~\ref{fig: COMs_static_collapse} present the temporal evolution of the abundances of selected COMs and their isotopic ratios. Radiolysis of ice components, such as CH$_4$, CH${_3}$OH, and H$_2$CO, produces suprathermal species (e.g., CH$_3$, CH$_2$OH, HCO), which undergo barrierless reactions with nearby species, rather than being saturated through hydrogenation.
These reactions dominate both the surface and mantle chemistry.
The $^{12}$C/$^{13}$C ratios of the produced COMs are governed by those of their precursor suprathermal species or radicals.
Also, the $^{12}$C/$^{13}$C ratios of these precursors reflect those of the parent ice species.
For instance, CH$_3$ radicals are formed from $^{13}$C-poor CH$_4$ and atomic C, while CH$_2$OH and CH$_3$O originate from $^{13}$C-rich CH$_{3}$OH and CO. 
Different carbon atoms within the same molecule can exhibit distinct isotope compositions when two precursors have different $^{12}$C/$^{13}$C ratios. {The isotope
ratios of selected COMs are listed in Table\ref{tab: COMs_ratio_static_collapse_vertical}.}

\begin{table*}[t!]
\centering
\caption{
Abundances and $^{12}$C/$^{13}$C ratios of selected COMs during the static and collapse phases.
For the static phase, values are taken at $10^{6}$ yr for bulk (surface + mantle) ice.
For the collapse phase, values correspond to gas after thermal desorption.
}
\label{tab: COMs_ratio_static_collapse_vertical}
\begin{tabular*}{\textwidth}{@{\extracolsep{\fill}} l l l l}
\hline\hline
\textbf{Species} & \textbf{Abundance ($n_i/n_\mathrm{H}$)} & \textbf{$^{12}$C/$^{13}$C (isotopomer 1)} & \textbf{$^{12}$C/$^{13}$C (isotopomer 2)} \\
\hline
\multicolumn{4}{l}{\textbf{Static phase at $10^{6}$ yr (bulk ice: surface + mantle)}} \\
\hline
HCOOCH$_3$ (ice)   & $1.2\times10^{-9}$  & 66 (H$^{13}$COOCH$_3$) & 63 (HCOO$^{13}$CH$_3$) \\
HOCH$_2$CHO (ice)  & $4.0\times10^{-10}$ & 66 (HOCH$_2$$^{13}$CHO) & 61 (HO$^{13}$CH$_2$CHO) \\
CH$_3$CHO (ice)    & $8.1\times10^{-10}$ & 97 ($^{13}$CH$_3$CHO) & 57 (CH$_3$$^{13}$CHO) \\
C$_2$H$_5$OH (ice) & $3.0\times10^{-10}$ & 34 ($^{13}$CH$_3$CH$_2$OH) & 32 (CH$_3$$^{13}$CH$_2$OH)\\
\hline
\multicolumn{4}{l}{\textbf{Collapse phase (gas after thermal desorption)}} \\
\hline
HCOOCH$_3$      & $1.3\times10^{-8}$ & 57 (H$^{13}$COOCH$_3$) & 57 (HCOO$^{13}$CH$_3$) \\
HOCH$_2$CHO     & $1.2\times10^{-7}$ & 57 (HOCH$_2$$^{13}$CHO) & 56 (HO$^{13}$CH$_2$CHO)\\
CH$_3$CHO       & $3.7\times10^{-9}$ & 79 ($^{13}$CH$_3$CHO) & 53 (CH$_3$$^{13}$CHO) \\
C$_2$H$_5$OH    & $6.1\times10^{-9}$ & 89 ($^{13}$CH$_3$CH$_2$OH) & 57 (CH$_3$$^{13}$CH$_2$OH) \\
\hline
\end{tabular*}
\end{table*}

\begin{figure*}[t!]
\centering
\includegraphics[width=0.47\textwidth]{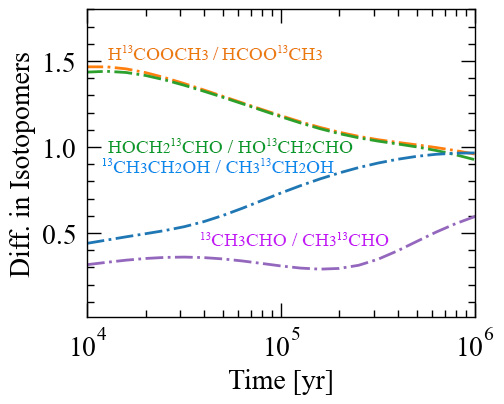}
\hfill
\includegraphics[width=0.47\textwidth]{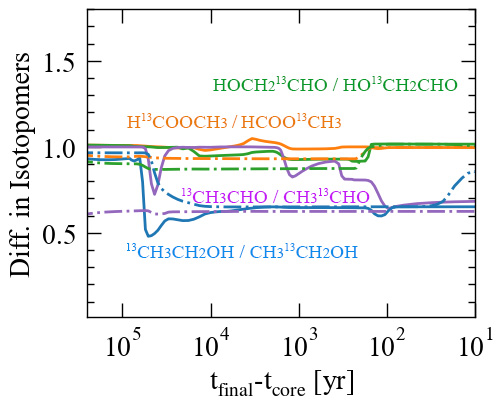}
\caption{
Temporal variation of the abundance ratio of isotopomers in the gas phase (solid lines) and in the bulk (surface + mantle) ice (dash-dot lines) during 
the static phase (left panel) and 
the collapse phase (right panel). 
The isotopomer ratios shown are H$^{13}$COOCH$_{3}$/HCOO$^{13}$CH$_{3}$, 
HOCH$_{2}$$^{13}$CHO/HO$^{13}$CH$_{2}$CHO, 
$^{13}$CH$_{3}$CHO/CH$_{3}$$^{13}$CHO, and 
$^{13}$CH$_{3}$CH$_{2}$OH/CH$_{3}$$^{13}$CH$_{2}$OH.
}
\label{fig: COMs_iso_ratio}
\end{figure*}

Figure~\ref{fig: COMs_iso_ratio} presents the abundance ratios of two $^{13}$C isotopomers (e.g., $^{13}$CH$_3$CHO/CH$_3^{13}$CHO), reflecting isotopomer-specific fractionation within functional groups. 
If the abundance ratio is unity, it means there is no difference in the abundances of the isotopomers.
Although the ratio for some COMs starts to deviate from unity around $10^4$ years, the ratio at later times ($\geq$10$^5$ yr) is more important as COMs abundances rise. Notably, CH$_3$CHO exhibits a persistent isotopomeric difference throughout the static phase, in contrast to other COMs where such differences are transient and diminish after $10^5$ years. This enduring distinction in CH$_3$CHO isotopomer ratio arises from its formation pathway: the methyl group is primarily derived from $^{13}$C-poor CH$_4$, whereas the formyl group (-CHO) comes from $^{13}$C-rich CO.

In contrast, ethanol (C$_2$H$_5$OH) does not retain isotopomer-specific fractionation at 10$^6$ years. Instead, both carbon atoms in the ethyl group become increasingly $^{13}$C-rich over time. Initially, CH$_3$ is derived from $^{13}$C-poor CH$_4$, and CH$_2$OH is formed from $^{13}$C-rich CH$_3$OH, yielding some isotopomer-specific fractionation. But as the contribution from the reaction between the C$_2$H$_5$ radicals and OH radicals—formed from symmetric $^{13}$C-rich C$_2$ (in turn originating from $^{13}$C-rich C$_3$)—becomes dominant, both carbons in the ethyl group inherit similar levels of $^{13}$C enrichment. This results in an overall $^{13}$C-rich ethyl group and the disappearance of isotopomeric differences.

Regarding methyl formate (HCOOCH$_3$) and glycolaldehyde (HOCH$_2$CHO), these COMs also show a transient isotopomeric difference, with a noticeable divergence appearing around $10^4$ years but diminishing by $10^5$ years. This is attributed to the combination of $^{13}$C-rich CH$_3$O and $^{13}$C-poor HCO radicals, which are derived from CH$_3$OH and H$_2$CO, respectively. The $^{13}$C depletion in H$_2$CO originates from the insertion of $^{13}$C-poor atomic carbon into H$_2$O ice, leading to the formation of H$_2$CO.
As the relative contribution of CH$_3$OH increases with time and the overall isotope ratios approach the ISM value, the isotopomer difference becomes less pronounced.

These results emphasize that isotopomer-resolved $^{12}$C/$^{13}$C ratios provide crucial insights not only into the molecular formation at each moment, but also into the longer-term chemical evolution and precursor supply that governs the isotopic structure of COMs during their formation.

\subsection{Collpase Phase}
Lower panels of Figure \ref{fig: COMs_static_collapse} show the resulting evolution in COMs abundances and their $^{12}$C/$^{13}$C ratios under the collapse phase.
When the dust temperature exceeds 20 K, thermal diffusion enables radical–radical reactions to occur efficiently on grain surfaces, potentially becoming a significant or even dominant pathway for COMs formation, depending on the mobility of reactants. This surface-induced chemistry gradually competes with, and in some cases supplements, COMs formation in the ice mantle.
This contrasts with the static phase, where COMs mainly form via suprathermal reactions in the ice mantle. When thermal diffusion dominates COMs formation, the carbon isotopic composition increasingly reflects the isotopic origins of individual surface radicals.
Among the radicals produced via photodissociation of major ice components, CH$_3$ radicals, which are $^{13}$C-poor due to their origin in CH$_4$ gas, play a particularly important role because their low binding energy (1600 K) enables efficient diffusion on grain surfaces even at $\sim$20 K. These mobile CH$_3$ radicals readily react with $^{13}$C-rich partners such as CH$_2$OH, CH$_3$O, and HCO, leading to the formation of COMs like C$_2$H$_5$OH. {The isotope
ratios of selected COMs are listed in Table\ref{tab: COMs_ratio_static_collapse_vertical}.
}


Figure~\ref{fig: COMs_iso_ratio} presents the abundance ratios of two $^{13}$C isotopomers.
C$_2$H$_5$OH serves as a representative example of isotopomer-specific evolution. At the end of the static phase, C$_2$H$_5$OH exhibits no significant isotopomer difference. However, after the protostar formation and CH$_3$ begins to diffuse and participate in surface chemistry, an isotopomer difference emerges. The methyl group in newly formed C$_2$H$_5$OH turns into $^{13}$C-poor due to its CH$_4$ gas origin, while the CH$_2$OH group is $^{13}$C-rich, reflecting the $^{13}$C-rich CH$_3$OH. This results in a divergence in $^{12}$C/$^{13}$C ratios between the two carbon positions, which is seen in the thermally desorbed C$_2$H$_5$OH.
The duration and efficiency of this diffusion-induced surface formation depend on the binding energy of the produced COMs. C$_2$H$_5$OH, with a relatively high binding energy (4680 K), continues to form and accumulate until temperatures approach $\sim$90 K. At that point, its abundance formed via surface diffusion exceeds that formed through non-diffusive processes in the ice mantle, and the collapse-phase chemistry largely determines its isotopic ratios.

In contrast, CH$_3$CHO maintains a persistent isotopomeric difference until the end of our simulations. The methyl group originates from $^{13}$C-poor CH$_4$ or atomic carbon, while the formyl group is formed from slightly $^{13}$C-rich CH$_3$OH. Surface formation of CH$_3$CHO remains limited during the collapse phase due to competing hydrogen abstraction reactions from CH$_3$CHO by other radicals (e.g., OH radical or H). Therefore, the isotopomeric ratio established during the static phase is almost preserved.

Other COMs, such as HCOOCH$_3$ and HOCH$_2$CHO, exhibit transient isotopomer differences. These species are primarily formed in the ice mantle from combinations of $^{13}$C-rich CH$_3$OH-derived radicals (e.g., CH$_3$O, CH$_2$OH, and HCO). However, in the collapse phase, their surface formation is suppressed due to CO desorption and the activation energy barrier of key reactions (e.g., CH$_3$O + CO $\rightarrow$ CH$_3$OCO + 2500 K) and the subsequent hydrogenation. Consequently, the isotopomer differences established in the static phase tend to persist in the mantle and are reflected in the thermally desorbed COMs.
Alternative radical–radical pathways involving HCO are possible, as HCO exhibits moderate mobility (the binding energy is 2280 K). However, its contribution is limited by competing hydrogen abstraction reactions involving atomic oxygen or other HCO radicals. As a result, surface production of formyl-bearing COMs remains low, and no new isotopomer difference is introduced.

Meanwhile, suprathermal-radical chemistry continues in the ice mantle, largely insulated from thermal diffusion processes inside the ice mantles due to the binding energies (5700 K) of species trapped in water ice. Here, radicals retain the $^{12}$C/$^{13}$C signatures established during the static phase. For example, both functional groups in HCOOCH$_3$ and HOCH$_2$CHO are formed from slightly $^{13}$C-rich CH$_3$OH and CO, resulting in little or no isotopomer difference.


When the temperature reaches the sublimation temperature of water ice ($T_{\rm sub}\sim 115$ K), these COMs are released into the gas phase together with water. 
{Here, $T_{\rm sub}$ is estimated by equating the thermal desorption rate with the accretion rate onto grains \citep{garrod2013three}.}
After sublimation, the isotopomer differences of COMs depend on both the mobility and isotopic composition of their precursor radicals, and on whether surface or mantle chemistry dominates the formation of each COM. 
If surface formation remains inefficient, the desorbed COMs will retain their mantle-like isotopic ratios, which are established in the static phase. 
This preservation emphasizes the importance of early chemical histories in shaping the isotopic characteristics of COMs in protostellar environments where ices have sublimated.

\section{Discussion} \label{sec: Discussion}
\subsection{Dependence on the Initial Condition of Carbon}
We assume that all carbon atoms are initially present as ionized carbon (C$^+$). To assess the impact of this assumption, we perform additional simulations throughout both the static and collapse phases, in which half of the initial carbon is present as CO, and the other half as C$^+$. In this case, the isotope exchange reaction $^{13}$C$^+$ + CO $\rightleftharpoons$ C$^+$ + $^{13}$CO + 35 K proceeds from the beginning of the simulation.
\begin{figure*}[t!]
\centering
\includegraphics[height=0.38\hsize]{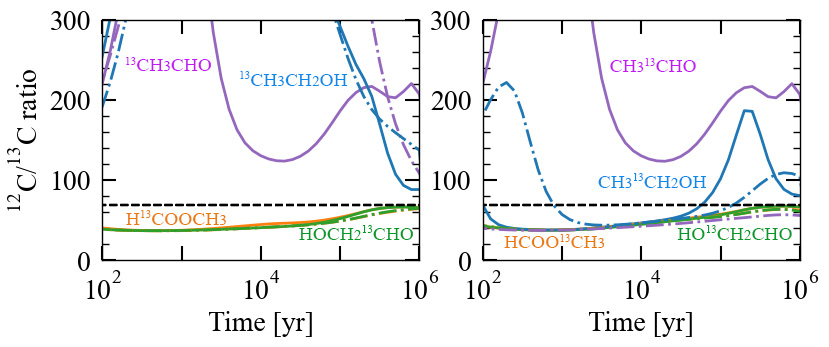}
\caption{Temporal variation of the $^{12}$C/$^{13}$C ratios of selected COMs in the gas phase (solid lines) and in the bulk (surface + mantle) ice (dash-dotted lines) during the static phase, in a model where half of the initial carbon is present as CO and the other half as C$^+$.
\label{fig: COMs_halfCO}}
\end{figure*}

Figure \ref{fig: COMs_halfCO} presents the resulting evolution in the $^{12}$C/$^{13}$C ratios of certain COMs during the static phase.
The influence of the initial carbon condition on the $^{12}$C/$^{13}$C ratios and isotopomer differences in COMs depends on their formation timescales. For COMs (e.g., HCOOCH$_3$) that primarily form after $\sim 10^5$ years—around the CO depletion timescale—the impact of the initial carbon condition is negligible. After this point, the efficiency of isotope exchange reactions decreases, and only a minor fractionation in gas-phase species is transferred to COMs via subsequent ice chemistry. As a result, the initial isotope fractionation of carbon-bearing species has little effect on their final $^{12}$C/$^{13}$C ratios or isotopomer distributions.

In contrast, COMs (e.g., C$_2$H$_5$OH) that partially form before $\sim 10^5$ years are significantly affected by the initial carbon condition. Because CO ice is still scarce prior to significant depletion, CO-based ice chemistry plays a limited role in early COM formation. As a result, early-forming COMs are mainly derived from gas-phase precursors that retain the imprint of early isotope fractionation processes. Consequently, their $^{12}$C/$^{13}$C ratios and isotopomer distributions reflect the initial C$^+$/CO ratio. For example, C$_2$H$_5$OH forms in the ice from C$_2$H$_5$, which in turn inherits the carbon isotope composition of gas-phase C$_3$. 
{When half of the carbon is initially in CO, the reduced abundance of atomic carbon suppresses the isotope exchange reactions between C and C$_3$, which would otherwise enrich C$_3$ in $^{13}$C.
Moreover, C$_3$ is mainly formed from the remaining atomic carbon with a relatively high $^{12}$C/$^{13}$C ratio, leading to an overall increase in the $^{12}$C/$^{13}$C ratio of C$_3$ and its derivatives \citep{2024ApJ...970...55I}.}
As a result, the $^{12}$C/$^{13}$C ratios in C$_2$H$_5$OH isotopomers in the ice phase increase, reaching values of $\sim$142 for C$_2$H$_5$OH/$^{13}$CH$_{3}$CH$_{2}$OH and $\sim$107 for C$_2$H$_5$OH/CH$_{3}$$^{13}$CH$_{2}$OH at $10^6$ years.
These elevated ratios are slightly preserved after thermal desorption during the collapse phase, particularly if thermal diffusion reactions are inefficient. For instance, the $^{12}$C/$^{13}$C ratios remain as high as $\sim$114 for C$_2$H$_5$OH/$^{13}$CH$_{3}$CH$_{2}$OH and $\sim$63 for C$_2$H$_5$OH/CH$_{3}$$^{13}$CH$_{2}$OH in the gas phase {after water ice sublimation}.
These findings highlight the importance of considering the initial carbon reservoir when interpreting carbon isotope ratios in COMs, particularly for species that form early in cold star-forming environments.

\subsection{Comparison to the Full-Scrambling Model}
In most of the previous astrochemical models, such as \citet{2024ApJ...970...55I}, the full-scrambling was assumed, and all carbon atoms within a molecule are treated as chemically equivalent. This approach simplifies the construction of a chemical network with $^{13}$C species but prohibits the study of the isotopomer-specific carbon isotope ratio of COMs. {In this model, isotopomer differences remain constant in time.}

In contrast, our position-conserved model tracks the position of each carbon atom in molecules through chemical reactions. This allows us to study the isotopomer-specific $^{12}$C/$^{13}$C ratios of COMs in star-forming cores using astrochemical kinetic models for the first time. In this framework, isotopomers of a given molecule can exhibit different abundances if their respective carbon atoms are derived from chemically distinct reservoirs, as mentioned in Section \ref{sec: results}

While we confirmed that the average of the $^{12}$C/$^{13}$C ratios \\
(e.g., 2$\times$[CH$_3$CHO]/([$^{13}$CH$_3$CHO]+[CH$_3$$^{13}$CHO])) is identical between the position-conserved model and the full-scrambling model, the latter model fails to capture the internal isotopomer distribution within molecules. Our results demonstrate that position-specific isotope modeling is essential for interpreting observations that resolve $^{13}$C substitutions at different molecular sites, and for identifying formation pathways of COMs based on their isotopic structure.

\subsection{Isotopomer Differences in COMs Governed by the $^{12}$C/$^{13}$C Ratios of Precursor Molecules}

\begin{figure*}[t!]
\centering
\includegraphics[width=0.47\textwidth]{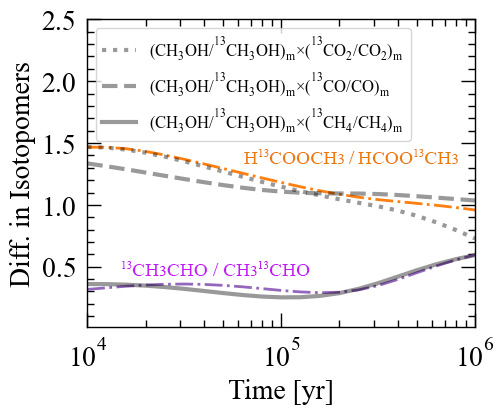}
\hfill
\includegraphics[width=0.47\textwidth]{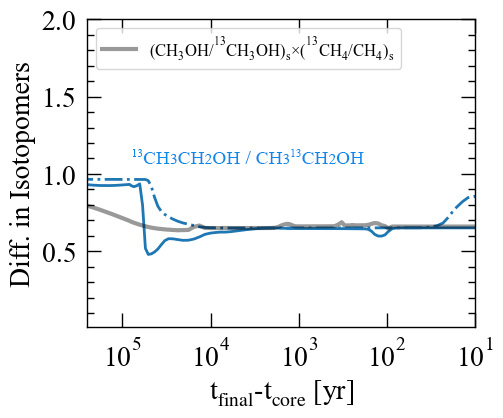}
\caption{
Temporal variation of the isotopomer differences during 
the static phase (left panel) and 
the collapse phase (right panel). 
In the left panel, the orange and purple lines represent HCOOCH$_3$ and CH$_3$CHO, respectively, which correspond to the same quantities as the left panel in Figure~\ref{fig: COMs_iso_ratio}. 
The dotted, dashed, and solid gray lines represent 
(CH$_3$OH/$^{13}$CH$_3$OH)$_m$$\times$($^{13}$CO$_2$/CO$_2$)$_m$, 
(CH$_3$OH/$^{13}$CH$_3$OH)$_m$$\times$($^{13}$CO/CO)$_m$, and 
(CH$_3$OH/$^{13}$CH$_3$OH)$_m$$\times$($^{13}$CH$_4$/CH$_4$)$_m$ in the ice mantle, respectively (``m" denotes the ice mantle).  
In the right panel, the solid and dash-dotted blue lines represent the isotopomer differences of C$_2$H$_5$OH in the gas phase and in the bulk (surface + mantle) ice, respectively, corresponding to the right panel in Figure~\ref{fig: COMs_iso_ratio}. 
The gray line represents 
(CH$_3$OH/$^{13}$CH$_3$OH)$_s$$\times$($^{13}$CH$_4$/CH$_4$)$_s$ 
in the ice surface (``s" denotes the ice surface).
}
\label{fig: COMs_13cmix}
\end{figure*}

In cold environments where thermal diffusion is inefficient, COMs primarily form within the ice mantle through non-thermal processes involving suprathermal species. In this regime, the $^{12}$C/$^{13}$C ratios of COMs and, in particular, their isotopomer-specific differences reflect the isotopic composition of their precursor radicals at the time of formation.

Figure~\ref{fig: COMs_13cmix} shows the abundance ratio of $^{13}$C isotopomers for CH$_3$CHO and HCOOCH$_3$ in the bulk ice during the static phase. These isotopomer ratios are compared against products of precursor isotope ratios such as (CH$_3$OH/$^{13}$CH$_3$OH)$_m$$\times$($^{13}$CO/CO)$_m$ and \\(CH$_3$OH/$^{13}$CH$_3$OH)$_m$$\times$($^{13}$CO$_2$/CO$_2$)$_m$, where the subscript $_m$ indicates the abundance ratio in the ice mantle. These products represent the expected isotopomer ratios based on the $^{12}$C/$^{13}$C ratios of precursor species involved in two-body formation reactions.
Before the depletion of gas-phase CO, $^{13}$C-poor atomic carbon contributes to CH$_3$OH formation via the ER reaction of gas-phase atomic carbon with H$_2$O ice and subsequent hydrogenation. Then CO$_2$ is more enriched in $^{13}$C than CH$_3$OH, although both molecules primarily are formed from CO. As a result, both (CH$_3$OH/$^{13}$CH$_3$OH)$_m$$\times$($^{13}$CO$_2$/CO$_2$)$_m$ and (CH$_3$OH/$^{13}$CH$_3$OH)$_m$$\times$($^{13}$CO/CO)$_m$ are different from unity. In this case, the isotopomer difference of COMs could be significant even though both reactants of COMs are enriched in $^{13}$C. However, when the hydrogenation of CO becomes the main formation pathway of CH$_3$OH due to the freeze-out of gas-phase CO, CO and CH$_3$OH have almost the same $^{12}$C/$^{13}$C ratio and 
(CH$_3$OH/$^{13}$CH$_3$OH)$_m$$\times$($^{13}$CO/CO)$_m$ get close to unity.

HCOOCH$_3$ is formed from only $^{13}$C-rich reactants. Before the freeze-out of gas-phase CO, CO in ice is mainly locked up as CO$_2$, so HCOOCH$_3$ is formed from the suprathermal CO, which is formed from CO$_2$ via radiolysis. As a result the isotopomer difference of HCOOCH$_3$ is overlapped on the (CH$_3$OH/$^{13}$CH$_3$OH)$_m$$\times$($^{13}$CO$_2$/CO$_2$)$_m$. After that, the formation of CH$_3$OH is dominated by the hydrogenation of CO, leading to HCOOCH$_3$ formation through CH$_3$OH and CO, and the isotopomer difference of HCOOCH$_3$ gets close to (CH$_3$OH/$^{13}$CH$_3$OH)$_m$$\times$($^{13}$CO/CO)$_m$.

CH$_3$CHO is formed from CH$_3$ and HCO radicals. While HCO is $^{13}$C-rich, CH$_3$ originates from both CH$_4$ and CH$_3$OH. In the early static phase, suprathermal CH$_3$ is mainly produced from CH$_4$, which is synthesized from $^{13}$C-poor atomic carbon. As a result, CH$_3$ retains a high $^{12}$C/$^{13}$C ratio. As the static phase progresses and CH$_3$OH abundance increases via hydrogenation of CO, radiolysis of CH$_3$OH begins to supply $^{13}$C-rich CH$_3$. The methyl group in CH$_3$CHO thus reflects a combination of these two sources. Consequently, the isotopomer ratio of CH$_3$CHO gradually approaches unity over time but remains lower than unity.


Thermal diffusion of radicals on grain surfaces also plays a role in setting the isotopomer ratios of COMs as well as radiolysis-induced two-body reactions in the ice mantle. Because radical–radical reactions proceed rapidly once the temperature reaches high enough for the diffusion of one of the reactants, the relative mobility of each radical may affect which formation pathways dominate. 
Figure~\ref{fig: COMs_13cmix} illustrates the abundance ratio of $^{13}$C isotopomers of C$_2$H$_5$OH together with the  $^{12}$C/$^{13}$C ratios of its precursors (CH$_4$ and CH$_3$OH) in the ice surface during the collapse phase. The isotopomeric difference directly traces the isotope compositions of CH$_3$ and CH$_2$OH radicals, which are primarily formed from $^{13}$C-poor CH$_4$ and $^{13}$C-rich CH$_3$OH, respectively. As such, the abundance ratio of $^{13}$CH$_{3}$CH$_{2}$OH/CH$_{3}$$^{13}$CH$_{2}$OH closely follows the (${^{13}}$CH$_3$OH/CH$_3$OH)$_s$$\times$(CH$_4$/${^{13}}$CH$_4$)$_s$, where the subscript $_s$ indicates the abundance ratio on the ice surface.

\subsection{Effects of Thermal Radical Diffusion} \label{subsec: BEofR}
\begin{figure}[ht!]
\centering
\includegraphics[height=1.3\hsize]{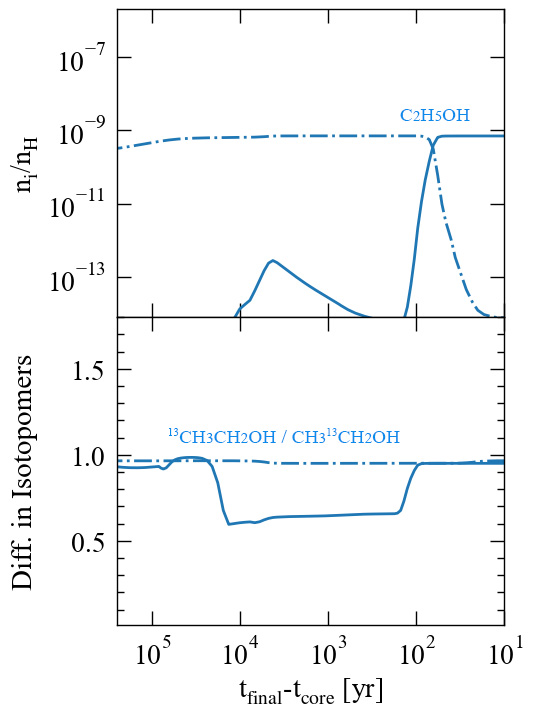}
\caption{Temporal variation of the abundance (upper panel) and isotopomer differences (lower panel) of C$_2$H$_5$OH gas (solid blue line) and bulk (surface + mantle) ice (dash-dotted blue line). A lower panel is similar to the right panel in Figure \ref{fig: COMs_13cmix}, but using $E_{\mathrm{diff}} = 0.6 \times E_{\mathrm{bind}}$ for selected radicals: CH$_3$, HCO, NH$_2$, and OH.
\label{fig: etha2}}
\end{figure}
The radical-radical reaction on the warm ice surface has little to no activation energy, making the reaction rate primarily controlled by the diffusion rate of the radical reactant \citep{2022ApJS..259...39E}. However, radicals are highly reactive, making it challenging to determine their diffusion activation energies accurately \citep{2022ApJ...940L...2M}. Additionally, the binding energy could also depend on the properties of the surface or the distribution of adsorption sites.
To examine how uncertainties in radical mobility affect isotopomer ratios, we varied the diffusion energy ($E_{\mathrm{diff}}$) from 0.4 to 0.6 times the binding energy ($E_{\mathrm{bind}}$) for key radicals: CH$_3$, HCO, NH$_2$, and OH. These species are crucial intermediates in COMs formation and are produced from abundant ice components.
Figure~\ref{fig: etha2} shows the impact of the radical mobility on the abundance and the isotopomer-specific $^{12}$C/$^{13}$C ratio of C$_2$H$_5$OH. The increased diffusion activation energies suppress the formation of C$_2$H$_5$OH on the surface, reducing the contribution of the diffusive radical-radical reactions on the surface to the overall formation of C$_2$H$_5$OH. Consequently, the isotopomer ratio $^{13}$CH$_{3}$CH$_{2}$OH/CH$_{3}$$^{13}$CH$_{2}$OH increases to 0.95 in the evaporated gas at the end of the simulation. Both isotopomers are enriched in $^{13}$C, with $^{12}$C/$^{13}$C ratios of 36 for $^{13}$CH$_{3}$CH$_{2}$OH and 35 for CH$_{3}$$^{13}$CH$_{2}$OH, respectively, where these values are comparable to those seen in the static phase.

These results underscore that isotopomer-resolved carbon isotope ratios reflect the formation pathway of COMs. When radical diffusion is efficient, surface formation pathways dominate, and the isotopic compositions of individual functional groups reflect the isotopic signatures of the radicals involved, especially CH$_3$, which often retains the $^{13}$C-poor character of its CH$_4$ precursor.
In contrast, when the activation energy of radical diffusion is high and diffusion is limited, the formation of COMs on the surface becomes inefficient. This enhances the relative contribution from the ice mantle, where suprathermal-radical reactions dominate, and the carbon isotope ratios of both positions often converge due to a common origin in dominant ice species (see Sect \ref{subsec: static}). As a result, the isotopomeric differences in the desorbed products diminish, and the $^{12}$C/$^{13}$C ratios approach mantle-like values, which are typically more $^{13}$C-enriched but less positionally differentiated.

Given the uncertainty in diffusion activation energies of radicals, isotopomer-specific $^{12}$C/$^{13}$C ratios of COMs may serve as a useful diagnostic for assessing the relative contributions of surface versus bulk chemistry in astrochemical environments.
{The binding and diffusion energies can vary depending on the physical properties of grain surfaces and the distribution of adsorption sites  \citep{2017SSRv..212....1C}.
{In recent years, significant efforts have been made to improve the accuracy of molecular binding energies, including quantum-chemical calculations on more realistic amorphous ice surfaces designed to constrain adsorption energetics \citep{2022ESC.....6.1514T, 2023ApJ...951...32T, 2024cosp...45.2296B, 2025MNRAS.539...82B, 2025A&A...698A.284G}. These studies highlight that binding energies may exhibit broad distributions rather than single values, underscoring the need for more physically realistic treatments in astrochemical modeling.
Despite these advances,} incorporating such macroscopic surface heterogeneity into current astrochemical models remains challenging, as the actual characteristics of interstellar dust—such as porosity, amorphous structure, and composition—are still poorly constrained.}
With the increasing availability of high-resolution isotopic observations, future laboratory measurements and quantum chemical calculations of radical diffusion and binding energies under realistic amorphous ice conditions will be essential for interpreting isotopic signatures in the context of COMs formation pathways.

\subsection{Comparisons with Observations}

We compare our modeling results with observational data from the ALMA-PILS survey of IRAS 16293-2422B \citep{2016Jorgensen,2018Jorgensen}. 
In these observations, the $^{12}$C/$^{13}$C ratios for the two $^{13}$C isotopomers of C$_2$H$_5$OH and CH$_3$CHO appear nearly identical. However, the reported uncertainties in the $^{12}$C/$^{13}$C ratio are as high as 30 \%, making it difficult to draw any definitive conclusions about isotopomeric differences. Such potential differences may be entirely obscured by the current measurement limitations.

In contrast, our model predicts a clear isotopomeric difference in the $^{12}$C/$^{13}$C ratios of thermally desorbed C$_2$H$_5$OH. This difference originates from the thermal diffusion reactions occurring during the collapse phase, which imprint distinct isotope signatures on the methyl and ethyl functional groups. However, if C$_2$H$_5$OH forms primarily in the ice mantle or under cold conditions, the isotopomeric difference is suppressed, as discussed in Section \ref{subsec: BEofR}. {In that case, the $^{12}$C/$^{13}$C ratio of C$_2$H$_5$OH can become as low as $\sim$35, consistent with the observed value of 41$\pm$12.
However, the static model alone cannot reproduce the observed ice abundance, even when radiolysis chemistry is included, suggesting that additional non-diffusive formation pathways, such as those proposed by \citet{2020ApJS..249...26J}, may be required to enhance ethanol production under cold conditions.}

Thermally desorbed CH$_3$CHO in our model also shows a persistent isotopomeric difference in both the static and collapse phases, again differing from the observations. 
However, this discrepancy may be resolved if a significant fraction of CH$_3$CHO originates from gas-phase carbon chain species with two indistinguishable carbon atoms, such as C$_2$H$_3$ and C$_2$H$_5$ {(see Section \ref{subsec: static})}. 
These species react with atomic oxygen in the gas phase, forming CH$_2$CO and CH$_3$CHO, respectively. Moreover, CH$_2$CO ice could form CH$_3$CHO ice via successive hydrogenation \citep{2022Fedoseev, ferrero2023formation, ibrahim2024significant}. 

\begin{figure*}[t!]
\centering
\includegraphics[height=0.27\hsize]{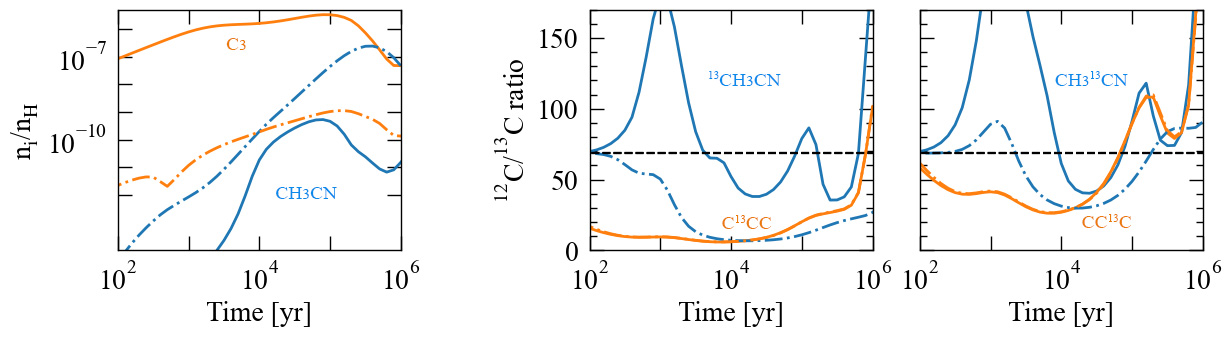}
\vspace{0.4cm} 
\includegraphics[height=0.278\hsize]{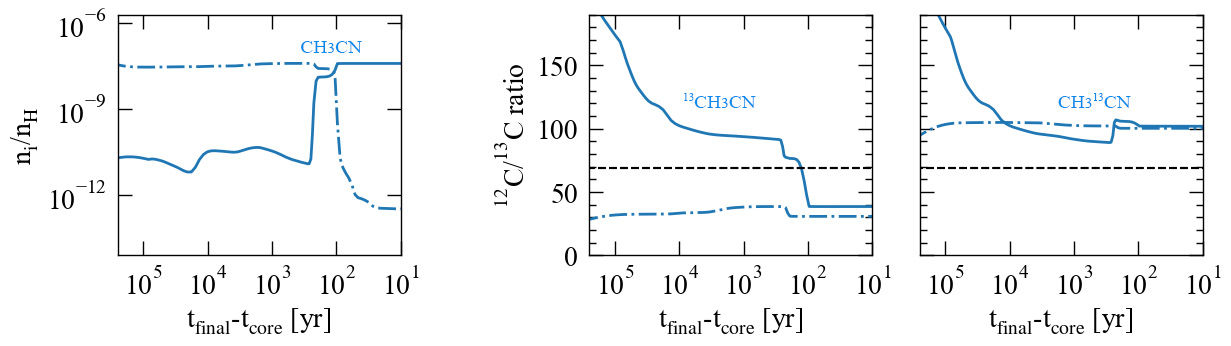}
\caption{
Temporal variation of the abundances (left panels) and $^{12}$C/$^{13}$C ratios (middle and right panels) of CH$_3$CN in the gas phase (solid lines) and bulk (surface + mantle) ice (dash-dotted lines) during the static phase (upper panels) and the collapse phase (lower panels). 
In the upper panels, the results for C$_3$ are also shown for comparison. 
The $^{12}$C/$^{13}$C ratios of C$^{13}$CC and CC$^{13}$C in the gas and ice phases are nearly identical, respectively.
}
\label{fig: CH3CN_static_collapse}
\end{figure*}

Additionally, we compare our results of $^{13}$CH$_3$CN/CH$_3$$^{13}$CN with that of the observations of Class 0/I protostellar systems obtained as part of the PRODIGE (PROtostars $\&$ DIsks: Global Evolution) large program \citep{2025A&A...699A.359B}. {The $^{12}$C/$^{13}$C ratios for the two $^{13}$C isotopomers of CH$_3$CN are lower than the ISM value in several Class 0/I sources, with some measurements—within their 1$\sigma$ uncertainties—falling below 40}. In contrast, our model predicts an isotopomer-specific fractionation in thermally desorbed CH$_3$CN: CH$_3$CN/$^{13}$CH$_3$CN $\sim$38 and CH$_3$CN/CH$_3$$^{13}$CN $\sim$102 (see lower panels in Figure \ref{fig: CH3CN_static_collapse}){, which reflect the $^{12}$C/$^{13}$C ratios of CH$_3$CN ice formed during the static phase.}
Upper panels in Figure \ref{fig: CH3CN_static_collapse} show the temporal evolution of the abundances and $^{12}$C/$^{13}$C ratios of CH$_3$CN in the gas and ice phases during the static phase. Prior to 10$^5$ years, the $^{13}$C isotopomers of CH$_3$CN ice are initially enriched in $^{13}$C, but their $^{12}$C/$^{13}$C ratios gradually increase over time. This trend reflects the formation pathway of CH$_3$CN ice, which originates from the dissociation of C$_2$H$_5$CN ice during the static phase.
The precursor molecule C$_2$H$_5$CN is formed via the reaction of atomic nitrogen with C$_3$, followed by successive hydrogenation on the ice. Consequently, the carbon isotope ratios in CH$_3$CN inherit those of C$_3$.
The $^{12}$C/$^{13}$C ratio of C$_3$ gas is primarily shaped by three isotope exchange reactions:
$^{13}$C + C$_3$ $\rightleftharpoons$ C + C$_2$$^{13}$C + 27 K, $^{13}$C + C$_3$ $\rightleftharpoons$ C + C$^{13}$CC + 43 K, and
C + C$_2$$^{13}$C $\rightleftharpoons$ C + C$^{13}$CC + 16 K.
In our model, C$_3$ is treated as a symmetric molecule, making $^{13}$CC$_2$ and C$_2$$^{13}$C indistinguishable.
As the efficiencies of the isotope exchange reactions decrease over time, C$_3$ increasingly forms from atomic carbon that is relatively $^{13}$C-poor. This results in a gradual increase in the $^{12}$C/$^{13}$C ratio of C$_3$. Furthermore, as more atomic carbon becomes locked in CO, the overall abundance of C$_3$ diminishes. This reduction in abundance amplifies the extent of isotope fractionation, leading to even higher $^{12}$C/$^{13}$C ratios and enhancing the differences between C$_3$ isotopomers.
This isotopic signature in the gas phase is subsequently transferred to the ice phase. {Since atomic nitrogen reacts with the terminal carbon atoms of C$_3$ in the ice phase}, the carbon atom adjacent to nitrogen in CH$_3$CN becomes more depleted in $^{13}$C than the methyl carbon. 

The discrepancy between our model and the observations may arise from physical processes during the static phase. A substantial {abundance} of CH$_3$CN ice ($\sim$4.4$\times$10$^{-8}$) is already formed by 10$^5$ years, and these molecules are enriched in $^{13}$C (CH$_3$CN/$^{13}$CH$_3$CN $\sim$10, CH$_3$CN/CH$_3$$^{13}$CN $\sim$46).
If the static phase effectively ends around 10$^5$ years, this $^{13}$C enrichment in CH$_3$CN ice may be preserved through to the protostellar stage, depending on subsequent thermal and chemical processing.
Another possible explanation is the uncertainty in the formation pathways of CH$_3$CN. While gas-phase reactions involving protonated CH$_3$CN or surface reactions between CH$_3$ and CN radicals may influence the $^{12}$C/$^{13}$C isotopomer ratios, as discussed by \citet{2025A&A...699A.359B} and \citet{2025A&A...699A.235E}, their actual contributions remain uncertain.

To better interpret isotopomer-resolved $^{12}$C/$^{13}$C ratios and constrain the formation pathways of COMs, further improvements in both observational precision and theoretical modeling are essential.
{As discussed at the beginning of this section, the observed $^{12}$C/$^{13}$C ratio of CH$_3$CHO is $\sim$67, with an uncertainty of about 30\% (i.e., 47–87).
Our results (CH$_3$CHO/$^{13}$CH$_3$CHO $\sim$80 and CH$_3$CHO/CH$_3$$^{13}$CHO $\sim$55) are consistent within these observational errors.}
{To evaluate the observational accuracy required to distinguish isotopomer differences, we estimated the detectable limit using both the relative percent difference (RPD) and error propagation analysis.
The RPD is defined as
\begin{equation}
    \mathrm{RPD} = \frac{|R_1 - R_2|}{(R_1 + R_2)/2} \times 100\%
\end{equation}
where $R_1$ and $R_2$ denote the respective $^{12}$C/$^{13}$C ratios of the two isotopomers. 
Assuming equal relative observational uncertainties for the two isotopomers, the RPD provides a first-order measure of how precise the observations must be to distinguish the isotopomers.
In our model, the largest isotopomer difference occurs for CH$_3$CHO or C$_2$H$_5$OH, with $R_1$/$R_2$ $\approx$ 0.65 ($R_1$ $\le$ $R_2$), corresponding to RPD $\approx$ 42\%.
This indicates that their distinction requires observational uncertainties smaller than roughly half of this value ($\lesssim$20\%).
}

\subsection{Chemically Heterogeneous Ice} \label{subsec:hetero_ice}

\begin{figure}[t!]
\centering
\includegraphics[height=0.75\hsize]{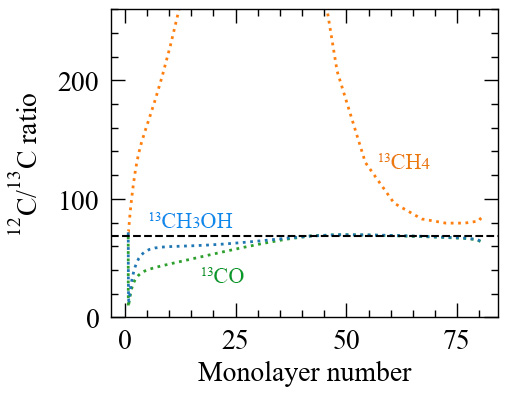}
\caption{Evolution of $^{12}$C/$^{13}$C ratios of selected species on the ice surface as a function of the cumulative number of icy layers. The “monolayer number” on the horizontal axis represents the temporal sequence of ice accumulation.
\label{fig: layer_static}}
\end{figure}

Figure \ref{fig: layer_static} shows the evolution of the $^{12}$C/$^{13}$C ratios of selected carbon-bearing species (e.g., CH$_3$OH, CO, CH$_4$) on the ice surface as a function of the number of monolayers accumulated over time. In the three-phase model adopted here, the surface layer is continuously updated as molecules accrete from the gas phase, while previously formed surface material is incorporated into the ice mantle. Note that the mantle phase in this model is not explicitly layered; thus, ``monolayer number" should be interpreted as a time-ordered index of ice growth.
These results demonstrate the formation of chemically heterogeneous ice, characterized by vertical stratification in carbon isotope composition.

The $^{12}$C/$^{13}$C ratios of surface species generally follow those in the gas phase, as ice growth proceeds through accretion from the gas. Because the gas-phase isotope ratio evolves rapidly, particularly during the early stages due to isotope exchange reactions, the newly accreted surface layers reflect the time-dependent gas-phase conditions. Once CO becomes the dominant carbon-bearing species, the efficiency of exchange reactions declines, resulting in diminished fractionation in the later-deposited surface layers.

In contrast, the ice mantle—composed of earlier surface layers buried over time—is less chemically active and gradually decouples from ongoing gas-phase processes (see dash-dotted lines in Figure \ref{fig: dominantC_static}). As a result, it preserves stronger signatures of the early isotope fractionation, producing a vertically stratified $^{12}$C/$^{13}$C pattern analogous to that seen in the deuteration study \citep{taquet2014multilayer} and in the nitrogen isotope fractionation study \citep{2018ApJ...857..105F}.

This vertical heterogeneity is also seen for species formed by surface hydrogenation, such as CH$_3$OH. While CH$_3$OH on the surface inherits the evolving isotope ratio of CO, the mantle CH$_3$OH retains a more $^{13}$C-enriched signature corresponding to early-time CO. Consequently, the carbon isotope ratios of CH$_3$OH in the mantle differ from those on the surface, even though they share a common formation pathway.

Radiolysis may also contribute to isotopic mixing in the ice mantle. For instance, radiolysis of CO via
\begin{equation}
    \rm CO \rightarrow C^* + O^*, \quad C^* \rightarrow C,
\end{equation}
can blend $^{13}$C-rich and $^{13}$C-poor components by converting CO into atomic carbon. However, this process is similar to photodissociation via UV photons induced by cosmic rays or external sources. Our results indicate that for dominant species such as CH$_4$ and CH$_3$OH, the effect of radiolysis on $^{12}$C/$^{13}$C ratios is minor, as these species primarily form via barrierless or low-barrier reactions independent of suprathermal chemistry.

This stratification in precursor isotope ratios suggests that COMs formed within the ice mantle may inherit analogous isotopic gradients, depending on their formation depths and the temporal evolution of precursors. 
While our model distinguishes between gas, surface, and mantle phases, it lacks a fully layered treatment of the mantle, and thus cannot capture intra-mantle isotopic variations in COMs formation. 
To quantitatively evaluate how such vertical isotope gradients influence the isotopomer-specific $^{12}$C/$^{13}$C ratios of COMs, future modeling efforts must incorporate a fully multilayered treatment of ice chemistry.

\section{Conclusions} \label{sec: sum}
We have developed a position-specific isotopic astrochemical model that traces the $^{12}$C/$^{13}$C ratios of individual carbon atoms within COMs in interstellar environments. By distinguishing $^{13}$C substitution sites, we reveal how isotopomer-specific carbon isotope ratios evolve from prestellar cores to protostellar cores. {This work extends our previous model \citep{2024ApJ...970...55I}, which included multi-carbon COMs but treated all carbon atoms as equivalent, to an isotopomer-specific isotopic model that distinguishes individual $^{13}$C substitution sites within each molecule.}
The main conclusions are as follows:

\begin{enumerate}

    \item {Isotopomer-Specific Carbon Isotope Ratio: COMs can exhibit up to 40\% differences in $^{12}$C/$^{13}$C ratios between their isotopomers, depending on the isotopic composition of their functional-group precursors. For example, CH$_3$CHO shows a persistent isotopomer difference between its $^{13}$CH$_3$CHO and CH$_3$$^{13}$CHO forms ($^{13}$CH$_3$CHO / CH$_3$$^{13}$CHO $\sim$0.6), originating from the distinct isotope ratios of CH$_4$ and CO in the early cold phase. These differences are absent in conventional models assuming complete atomic scrambling.}

    \item Cold-Phase Isotopic Memory:
    In the static phase, COMs form primarily in the ice mantle via non-thermal and suprathermal reactions in our model. The isotopomeric compositions of COMs reflect the $^{12}$C/$^{13}$C ratios of precursors like CH$_3$OH, CO, and CH$_4$, which themselves inherit isotope fractionation patterns from earlier gas-phase isotope exchange reactions. This imprint can persist throughout the evolution and remains in thermally desorbed COMs if surface formation is inefficient.

    \item {Collapse-Phase Surface Chemistry: Diffusion-induced surface chemistry produces new isotopomer differences not present at the end of the static phase. For instance, as dust grains warm, $^{13}$C-poor CH$_3$ radicals diffuse and react with $^{13}$C-rich radicals such as CH$_2$OH, forming $^{13}$C$_2$H$_5$OH. Consequently, the $^{13}$CH$_3$CH$_2$OH / CH$_3$$^{13}$CH$_2$OH ratio in ice decreases from unity to $\sim$0.6. This demonstrates that isotopomer differences can be amplified during warm-up through selective surface diffusion.}

    \item {Sensitivity to Diffusion Activation Energy Barriers:
    A 20\% increase in diffusion activation energies reduces the contribution of surface formation and isotopomer differences. For example, $^{13}$CH$_3$CH$_2$OH / CH$_3$$^{13}$CH$_2$OH in ice does not change during the collapse phase.  
    Under these conditions, thermally desorbed COMs more strongly reflect the mantle composition, highlighting the need for accurate estimates of radical mobility.}

    \item {Comparison with Observations:
    Our model reproduces the observed $^{12}$C/$^{13}$C ratios of CH$_3$CHO within the measurement uncertainty ($\sim$67$\pm20$, i.e., 47--87), with modeled values of $\sim$80 and $\sim$55 for the two isotopomers. 
    For C$_2$H$_5$OH, diffusion-driven surface chemistry during the collapse phase produces isotopomeric differences, whereas formation mainly in the ice mantle or under cold conditions suppresses them, yielding a ratio ($\sim$40) consistent with the observed value ($41$$\pm$12, i.e., 28--53). 
    In contrast, CH$_3$CN shows isotopomer-specific ratios (38 and 102) that differ from the $^{13}$C-enriched isotopomers observed in the PRODIGE survey, likely due to the preservation of $^{13}$C-rich CH$_3$CN ice formed during the static phase or the contribution of gas-phase reactions. 
    Future high-sensitivity isotopomer measurements—achieving uncertainties smaller than $\approx$20\% in the derived $^{12}$C/$^{13}$C ratios—will be required to distinguish between surface- and mantle-driven formation routes of COMs.}

    \item {Implications and Future Work:
    This position-specific isotopic modeling demonstrates that isotopomer-resolved $^{12}$C/$^{13}$C ratios can preserve information on the chemical origin of COMs from the cold to the warm stages of star formation. 
    Quantitative comparison with our previous model \citep{2024ApJ...970...55I}, which does not distinguish positions of carbon isotopes within a molecule, confirms that including isotopomer-specific chemistry refines the distribution of $^{13}$C among functional groups without altering the total carbon budget. 
    Further improvements in observational precision and constraints on microphysics on ice chemistry, such as diffusion barriers and ice morphology, will be essential to confirm isotopomer-specific signatures and link isotopic patterns across simple and complex molecules for understanding their formation pathways.}

\end{enumerate}

Our study demonstrates that the position-specific carbon isotope ratio of COMs retains a history of their formation processes and environmental effects. By resolving isotopomer differences, we gain access to chemical information that would otherwise be averaged out in conventional isotopic models. This framework offers a promising pathway toward decoding the formation pathways of complex organic molecules in star-forming regions.


\noindent
\textbf{Acknowledgement}: We thank the anonymous referees for helpful suggestions on the manuscript.
This work was supported by JSPS KAKENHI
(grant Nos. JP25KJ1333, 19K03910 and 25K07364).
TJM thanks the STFC for support through grant number ST/T000198/1.

\noindent
\textbf{Conflict of Interest}: The authors declare that the research was conducted in the absence of any commercial or financial relationships that could be construed as a potential conflict of interest.






\bibliographystyle{aasjournal}
\bibliography{achemso-demo}


\end{document}